\documentclass[prd,twocolumn,floatfix,superscriptaddress,showpacs,letterpaper,altaffilletter,nofootinbib]{revtex4}
\usepackage{color}
\usepackage{times}
\usepackage{graphicx}
\usepackage{fancyhdr}
\usepackage{float}
\usepackage{ulem}
\makeatletter
\makeatletter
\def\@fnsymbol#1{\ifcase#1\or * \or  $+$ \or  \$ \or \#  \or \dag \or \ddag \or
$\mathsection$ \or $ \mathparagraph$ \or $\|$  \or \textordfeminine \or \textbullet   
\or ** \or $++$ \or  \$\$ \or \#\#  \or \dag\dag \or \ddag\ddag \or
$\mathsection\mathsection$ \or $ \mathparagraph\mathparagraph$ \or $\|\|$  \or 
\textordfeminine\textordfeminine \or \textbullet \textbullet \or *** \or $+++$ 
\or  \$\$\$ \or \#\#  \or \dag\dag \or \ddag\ddag \or
$\mathsection \mathsection\mathsection$ \or $ \mathparagraph 
\mathparagraph\mathparagraph$ \or $\|\|\|$  \or 
\textordfeminine\textordfeminine\textordfeminine \or 
\textbullet\textbullet\textbullet \or \else \@ctrerr\fi}
\makeatother

\def\thercsid{\relax}
\def\rcsid#1{\def\next##1#1{\def\thercsid{##1}}\next}
\rcsid$Id: s2-macho-paper.tex,v 1.37 2005/05/10 02:08:13 duncan Exp $

\renewcommand{\today}{\number\day\space\ifcase\month\or
  January\or February\or March\or April\or May\or June\or
  July\or August\or September\or October\or November\or December\fi
  \space\number\year}

\begin{document}

\title{Search for Gravitational Waves from Primordial Black Hole Binary
Coalescences in the Galactic Halo
}

\date[\relax]{ RCS \thercsid; compiled \today }
\pacs{95.85.Sz, 04.80.Nn, 07.05.Kf, 97.80.--d}

\begin{abstract}
\quad
We use data from the second science run of the LIGO gravitational-wave
detectors to search for the gravitational waves from primordial black hole
(PBH) binary coalescence with component masses in the range
$0.2$--$1.0\,M_\odot$.  The analysis requires a signal to be found in the data
from both LIGO observatories, according to a set of coincidence criteria. No
inspiral signals were found. Assuming a spherical halo with core radius
$5$~kpc extending to $50$~kpc containing non-spinning black holes with masses
in the range $0.2$--$1.0\,M_\odot$, we place an observational upper limit on
the rate of PBH coalescence of $63$ per year per Milky Way halo (MWH) with
$90\%$ confidence.
\end{abstract}

%
%
%
\newcommand*{\AG}{Albert-Einstein-Institut, Max-Planck-Institut f\"ur Gravitationsphysik, D-14476 Golm, Germany}
\affiliation{\AG}
\newcommand*{\AH}{Albert-Einstein-Institut, Max-Planck-Institut f\"ur Gravitationsphysik, D-30167 Hannover, Germany}
\affiliation{\AH}
\newcommand*{\AN}{Australian National University, Canberra, 0200, Australia}
\affiliation{\AN}
\newcommand*{\CH}{California Institute of Technology, Pasadena, CA  91125, USA}
\affiliation{\CH}
\newcommand*{\DO}{California State University Dominguez Hills, Carson, CA  90747, USA}
\affiliation{\DO}
\newcommand*{\CA}{Caltech-CaRT, Pasadena, CA  91125, USA}
\affiliation{\CA}
\newcommand*{\CU}{Cardiff University, Cardiff, CF2 3YB, United Kingdom}
\affiliation{\CU}
\newcommand*{\CL}{Carleton College, Northfield, MN  55057, USA}
\affiliation{\CL}
\newcommand*{\FN}{Fermi National Accelerator Laboratory, Batavia, IL  60510, USA}
\affiliation{\FN}
\newcommand*{\HC}{Hobart and William Smith Colleges, Geneva, NY  14456, USA}
\affiliation{\HC}
\newcommand*{\IU}{Inter-University Centre for Astronomy  and Astrophysics, Pune - 411007, India}
\affiliation{\IU}
\newcommand*{\CT}{LIGO - California Institute of Technology, Pasadena, CA  91125, USA}
\affiliation{\CT}
\newcommand*{\LM}{LIGO - Massachusetts Institute of Technology, Cambridge, MA 02139, USA}
\affiliation{\LM}
\newcommand*{\LO}{LIGO Hanford Observatory, Richland, WA  99352, USA}
\affiliation{\LO}
\newcommand*{\LV}{LIGO Livingston Observatory, Livingston, LA  70754, USA}
\affiliation{\LV}
\newcommand*{\LU}{Louisiana State University, Baton Rouge, LA  70803, USA}
\affiliation{\LU}
\newcommand*{\LE}{Louisiana Tech University, Ruston, LA  71272, USA}
\affiliation{\LE}
\newcommand*{\LL}{Loyola University, New Orleans, LA 70118, USA}
\affiliation{\LL}
\newcommand*{\MP}{Max Planck Institut f\"ur Quantenoptik, D-85748, Garching, Germany}
\affiliation{\MP}
\newcommand*{\MS}{Moscow State University, Moscow, 119992, Russia}
\affiliation{\MS}
\newcommand*{\ND}{NASA/Goddard Space Flight Center, Greenbelt, MD  20771, USA}
\affiliation{\ND}
\newcommand*{\NA}{National Astronomical Observatory of Japan, Tokyo  181-8588, Japan}
\affiliation{\NA}
\newcommand*{\NO}{Northwestern University, Evanston, IL  60208, USA}
\affiliation{\NO}
\newcommand*{\SC}{Salish Kootenai College, Pablo, MT  59855, USA}
\affiliation{\SC}
\newcommand*{\SE}{Southeastern Louisiana University, Hammond, LA  70402, USA}
\affiliation{\SE}
\newcommand*{\SA}{Stanford University, Stanford, CA  94305, USA}
\affiliation{\SA}
\newcommand*{\SR}{Syracuse University, Syracuse, NY  13244, USA}
\affiliation{\SR}
\newcommand*{\PU}{The Pennsylvania State University, University Park, PA  16802, USA}
\affiliation{\PU}
\newcommand*{\TC}{The University of Texas at Brownsville and Texas Southmost College, Brownsville, TX  78520, USA}
\affiliation{\TC}
\newcommand*{\TR}{Trinity University, San Antonio, TX  78212, USA}
\affiliation{\TR}
\newcommand*{\HU}{Universit{\"a}t Hannover, D-30167 Hannover, Germany}
\affiliation{\HU}
\newcommand*{\BB}{Universitat de les Illes Balears, E-07122 Palma de Mallorca, Spain}
\affiliation{\BB}
\newcommand*{\BR}{University of Birmingham, Birmingham, B15 2TT, United Kingdom}
\affiliation{\BR}
\newcommand*{\FA}{University of Florida, Gainesville, FL  32611, USA}
\affiliation{\FA}
\newcommand*{\GU}{University of Glasgow, Glasgow, G12 8QQ, United Kingdom}
\affiliation{\GU}
\newcommand*{\MU}{University of Michigan, Ann Arbor, MI  48109, USA}
\affiliation{\MU}
\newcommand*{\OU}{University of Oregon, Eugene, OR  97403, USA}
\affiliation{\OU}
\newcommand*{\RO}{University of Rochester, Rochester, NY  14627, USA}
\affiliation{\RO}
\newcommand*{\UW}{University of Wisconsin-Milwaukee, Milwaukee, WI  53201, USA}
\affiliation{\UW}
\newcommand*{\WU}{Washington State University, Pullman, WA 99164, USA}
\affiliation{\WU}

\author{B.~Abbott}    \affiliation{\CT}
\author{R.~Abbott}    \affiliation{\LV}
\author{R.~Adhikari}    \affiliation{\LM}
\author{A.~Ageev}    \affiliation{\MS}  \affiliation{\SR}
\author{B.~Allen}    \affiliation{\UW}
\author{R.~Amin}    \affiliation{\FA}
\author{S.~B.~Anderson}    \affiliation{\CT}
\author{W.~G.~Anderson}    \affiliation{\TC}
\author{M.~Araya}    \affiliation{\CT}
\author{H.~Armandula}    \affiliation{\CT}
% ...

 \collaboration{The LIGO Scientific Collaboration, http://www.ligo.org}
 \noaffiliation
\maketitle

Gravitational waves from binary inspiral are among the most promising
sources for the first generation of gravitational wave interferometers. Data
from the first and second LIGO science runs has been searched for binary
neutron star coalescence with component masses in the range
$1$--$3\,M_\odot$~\cite{LIGOS1iul,LIGOS2iul}, and a search for binary black
holes with component masses $> 3\,M_\odot$ is underway~\cite{LIGOS2bbh}. 
Here we consider binaries with component masses in the range
$0.2$--$1\,M_\odot$. Such binaries must contain a pair of black holes in order
to be detectable by LIGO. Binaries composed of low mass stellar remnants, such
as white dwarfs, will coalesce before the gravitational waves from
inspiral reach a high enough frequency to be detected by ground based
interferometers\cite{thorne.k:1987}.  Black holes with masses $< 1\,M_\odot$
are assumed to be primordial black holes (PBHs) since there is no known
mechanism that can produce sub-solar mass black holes as a product of stellar
evolution.

There is evidence from gravitational microlensing surveys of the Large
Magellanic Cloud (LMC) that  $\sim 20\%$ of the Galactic halo is composed
of massive compact halo objects (MACHOs) with masses
$0.15$--$0.9\,M_\odot$~\cite{Alcock:2000ph}. At present the explanation of the
observed excess of microlensing events is controversial. Self lensing of stars
in the LMC cannot account for all the observed microlensing
events~\cite{Mancini:2004pb} and there are a number of potential problems with
all the events being due to white dwarfs in the
halo~\cite{Spagna:2004bg,Garcia-Berro:2004he}. The nature of the majority of
observed lenses is unknown~\cite{Bennett:2005af} and PBHs with masses $\sim
0.5\,M_\odot$ have been proposed as possible MACHO
candidates~\cite{Jedamzik:1996mr,Yokoyama:1999xi}.  If the MACHOs are PBHs, it
will be very difficult to determine this using electromagnetic
observations~\cite{Nakamura:1998ai}.
If such PBHs formed in the early universe, then it has been suggested that
some fraction of the PBHs may exist in binaries which are coalescing
today~\cite{Finn:1996dd,Nakamura:1997sm}.  If a significant fraction of
MACHOs are in the form of PBHs, then estimates of the rate of PBH binary
coalescence suggest that it may be a factor of $100$ greater than that of
binary neutron stars~\cite{Nakamura:1997sm,Ioka:1998nz}.  If this scenario is
correct, the PBH binaries are a promising source of gravitational waves and
the presence of PBHs in the halos of galaxies can be confirmed by the
detection of their coalescence.

In this paper we report on a search for PBH binaries in data
from the second LIGO science run (S2). The data analysis techniques used are
identical to those used to search for binary neutron stars in the S2
data~\cite{LIGOS2iul}, the only difference being in the choice of the search
parameters. No inspiral signals were found and so we place an upper limit on the
rate of PBH binary coalescence in the Galactic halo. We compare this observed
rate to that estimated from microlensing observations using the model of PBH
binary formation proposed in~\cite{Ioka:1998nz}. Finally we comment on possible
future rate limits as the LIGO detectors improve towards their design
sensitivity.

Data for the second science run was taken over $59$~days from February 14 to
April 14, 2003. All three LIGO detectors at the two observatories were
operational: a $4$~km and a $2$~km interferometer at the LIGO Hanford
Observatory (LHO), Washington, and a $4$~km interferometer at the LIGO
Livingston Observatory (LLO), Louisiana. These detectors are referred to as
H1, H2 and L1 respectively.  During operation, feedback to the mirror
positions and to the laser frequency keeps the optical cavities near
resonance, so that interference in the light from the two arms recombining at
the beam splitter is strongly dependent on the difference between the lengths
of the two arms.  A photodiode at the antisymmetric port of the detector
senses this light, and a digitized signal is recorded at a sampling rate of
16384~Hz.  This channel can then be searched for a gravitational wave signal.
More details on the detectors' instrumental configuration and performance can
be found in \cite{LIGOS1instpaper} and \cite{LIGOS2grb}.  In order to avoid
the possibility of correlated noise sources between the H1 and H2 detectors,
we only analyze data from times when the L1 detector is operational.  
We demand that a candidate event be coincident between the L1 and
one or both of the Hanford detectors to reduce the rate of background events
due to non-astrophysical sources. 

We refer to \cite{LIGOS2iul} for a detailed description of the data analysis
pipeline.  Briefly, we used matched filtering with a bank of filters
constructed using second order restricted post-Newtonian
templates~\cite{findchirp,LALS2MACHO}. The bank is designed so that the loss in
signal-to-noise ratio (SNR) between a putative signal and its nearest template
is no more than $5\%$~\cite{Owen:1998dk}. Data was filtered in $2048$~s chunks
and times when the SNR $\rho$ of a template exceeded a threshold $\rho > \rho^\ast$
were considered candidate triggers.  Noise transients in the data may yield
high values of SNR, so a time-frequency $\chi^2$ veto~\cite{Allen:2004} is
used to distinguish between such events and inspiral signals. The
computational resources required to perform the search are proportional to the
number of templates $N$, which scales as $N \sim m_\mathrm{min}^{-8/3}$, where
$m_\mathrm{min}$ is the smallest binary component mass in the bank. The
available resources limited the template bank to binaries with component
masses above $0.2\,M_\odot$. The number of templates fluctuates over the
course of the run due to changing detector noise, the average value being
$14\,178$ templates in the most sensitive detector (L1).  The low frequency
cut off of the search was $100$~Hz due to detector noise at lower frequencies;
the resulting template durations were between $4$ and $56$~s.  We can
determine the time of an inspiral to within $1$~ms, so to be considered
coincident, triggers must be observed within a time window $\delta t = 11$~ms
between LHO and LLO (the light travel time between the observatories is
$10$~ms) and within $\delta t = 1$~ms between the detectors at LHO.  We use
the template bank from the most sensitive detector for all three detectors and
demand that the mass parameters of coincident triggers are identical $\delta m
= 0$.  We demand that triggers in the LHO detectors pass an amplitude
consistency test. No amplitude test is applied to triggers from different
obervatories as the different alignment of LLO and LHO (due to their different
latitudes) can occasionally cause large variations in the detected signal
amplitudes for astrophysical signals. Many templates may be trigged nearly
simultaneously, forming clusters of triggers; the trigger with the largest SNR
from each cluster is chosen for further study; triggers separated by more than
$4$~s are considered unique.

The sensitivity of the detectors is measured by determining the maximum
distance to which the detector is sensitive to the inspiral of a pair
of $0.5\,M_\odot$ PBHs at SNR of 8; that is, the distance at which an
\emph{optimally oriented} binary would produce an SNR of 8. This distance is
refered to as the range of the detector. The detectors were
at differing stages of progress towards design sensitivity during the S2 run
and the sensitivity of each detector fluctuated over the course of the run in
response to different noise sources. The average range of the detectors during
S2 was $704$~kpc for L1, $359$~kpc for H1 and $239$~kpc for H2. As we demand
coincidence with the less sensitive Hanford detectors, the range of the search
is limited to the neighborhood of the Milky Way, although there are times when
L1 is sensitive to M31.  The PBH binary search uses the triggered search
pipeline described in \cite{LIGOS2iul} which takes advantage of coincidence
and the difference in detector sensitivity to reduce computational cost. Data
from the less sensitive detectors (H1 and H2) is only filtered if a trigger is
observed in the most sensitive detector (L1). Since we demand $\delta m = 0$,
the triggered search is functionally equivalent to filtering all three
detectors with the same template bank and looking for coincidence.

\begin{figure}
\begin{center}
\includegraphics[width=\linewidth]{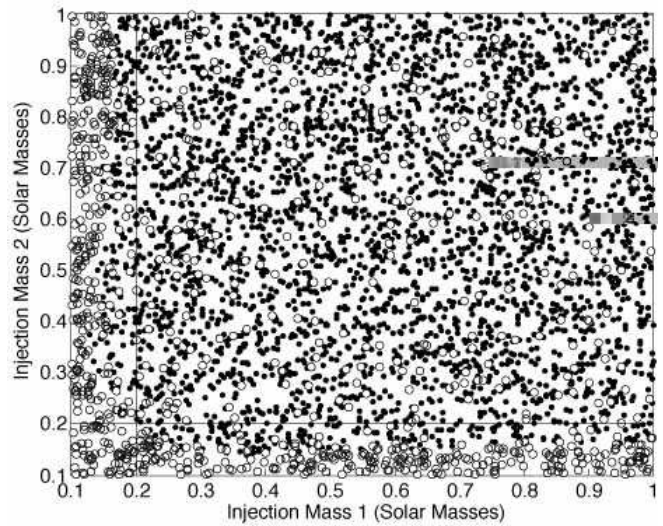}
\end{center}
\caption{\label{f:found_missed_inj}%
3270 inspiral signals were injected into the data using a uniform distribution
in $m_1$ and $m_2$. Each injection found by the pipeline is shown with a point
and each injection missed is shown by a circle. The lines at $m_1 = m_2 =
0.2\,M_\odot$ show the edge of the template bank. When constructing the upper
limit on the rate, we only consider injections that lie inside the region of
parameter space covered by the template bank, i.e. $m_1,m_2>0.2\,M_\odot$. It
can be seen that the sensitivity of the search is greatly reduced for binaries
with a component of mass $< \sim 0.15$.}
\end{figure}

We algorithmically select a subset (approximately $10\%$) of the data to be
used as \emph{playground} for tuning the analysis pipeline. The playground
samples the entire data set so that it is representative of the S2 data and
allows us to tune our data analysis pipeline without introducing statistical
bias into the upper limit. The goal of tuning the pipeline is to maximize the
efficiency of the pipeline to detection of gravitational waves from binary
inspirals without producing an excessive rate of spurious candidate events.
The false alarm rate of our search was set by the available computational
resources. If a trigger exceeds the SNR threshold, then a $\chi^2$ veto must
be performed at fifteen times the computational cost of a matched filter. This
limits the SNR threshold to $\rho_\ast = 7$ in all three detectors. We tune
our detection pipeline by attempting to maximize the detection efficiency of a
population of signals which are added to the data and then sought. Since we
are interested in PBH binaries in the halo of the Galaxy, the population we
inject is distributed as a standard spherical halo with density distribution
\begin{equation}
\rho(r) \propto \frac{1}{r^2 + a^2}
\label{e:halomodel}
\end{equation}
where $r$ is the Galactocentric radius and $a = 5$~kpc is the halo core
radius. The halo is truncated at $r = 50$~kpc. The component masses of the
binaries are uniformly distributed between $0.1$ and $1.0\,M_\odot$. Although
the template bank is terminated at a lower mass of $0.2\,M_\odot$, we were
able to tune the search so that it is possible to detect inspirals with component
masses down to $\sim 0.15\,M_\odot$, as shown in Fig~\ref{f:found_missed_inj}. 
Detection efficiencies for $m_1,m_2 \ge 0.2\,M_\odot$ were found to be greater
than $90\%$, consistent with that expected from consideration of the detector
sensitivities. We investigated injected signals whose masses were inside the
template parameter space but which were not recovered and found that the loss
was due to unfavorable alignment of the binary with the detector antenna
patterns. However, detection efficiency is uniform across the region of mass
parameter space covered by the template bank. For PBH binaries, the $\chi^2$
veto was found to be particularly powerful. No triggers were observed in the
playground data.

We estimate the background rate for this search by introducing an artificial
time offset $\Delta t \ge 17$~s to the triggers coming from the Livingston
detector relative to the Hanford detectors. We assume the shifted triggers are
uncorrelated between the observatories, however we do not time-shift the two
Hanford detectors relative to each other as there may be real correlations due
to environmental disturbances. The triggers which emerge at the end of the
pipeline are considered a single trial representative of the output from a
search if no signals are present in the data. By choosing a time shift much
greater than the light travel time between the observatories, we ensure that a
true gravitational wave signal will not be coincident in the time shifted
data. If the times of background triggers are uncorrelated between
observatories then the background rate is entirely due to accidental
coincidences which can be estimated using the time-shift analysis. A total of
$60$ time shifts were analyzed to estimate the background with $\Delta t = \pm
17 + 10n$~s, where $n = 0,1,\ldots,29$. 

\begin{figure}
\begin{center}
\includegraphics[width=\linewidth]{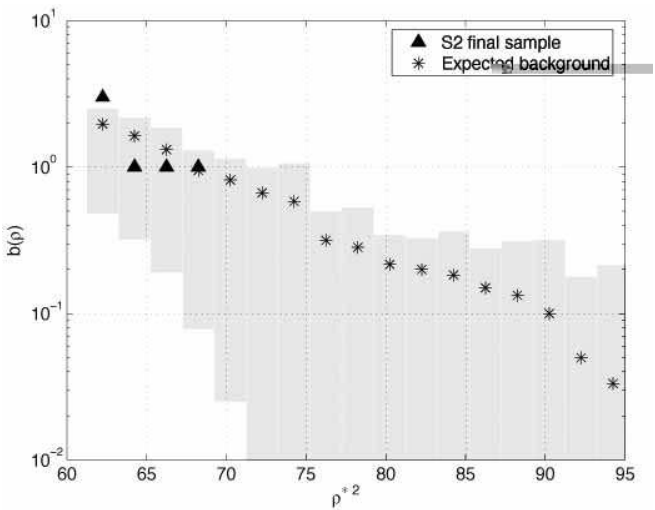}
\end{center}
\caption{\label{f:event-per-s2}%
The mean number of triggers per S2 observation time above combined SNR
$\rho^*$.  The stars represent the expected background based on 60 time shift
analyses.  The shaded envelope indicates the standard deviation in the number of
events.  The triangles show the distribution of events from the final S2
sample.}
\end{figure}

For a coincident trigger, the SNR observed in L1 is denoted $\rho_\mathrm{L}$
and the coherent SNR observed in H1 and H2 is $\rho_\mathrm{H}$. The
distribution of background triggers in the $(\rho_\mathrm{L},\rho_\mathrm{H})$
plane for the PBH binary search showed a similar distribution to that of the
binary neutron star search~\cite{LIGOS2iul}; the SNR of background triggers in
the Hanford detectors was typically larger than that in the Livingston
detector. In order to combine triggers from the two detectors, the
SNRs of the triggers were combined as
\begin{equation}
\rho^2 = \rho_\mathrm{L}^2 + \rho_\mathrm{H}^2/4
\label{e:combined-snr}
\end{equation}
with any coincident triggers in the Hanford detectors combined
coherently~\cite{LIGOS2iul,Pai:2000zt}. Fig~\ref{f:event-per-s2} shows the
sample mean and standard deviation of the expected number of accidental
coincidence events per S2 observation time with combined SNR $\rho^2 >
\rho^2_\ast$ computed from the 60 time shifts. This can be compared with the
triggers observed by the search to give a visual estimate of the significance
of the event candidates.  

The pipeline described above was used to analyze the S2 data.  After applying
the data quality cuts and the L1 instrumental veto described in
\cite{LIGOS2iul}, and discarding science segments with durations less than
$2048$~s, a total of $375$~hours of data was searched for binary coalescence.
For the upper limit analysis, we only considered the non-playground times
amounting to $341$~hours (The extra $2$~hours of data in this analysis
compared with~\cite{LIGOS2iul} is due to a bug fix applied to the data
handling routines after completion of the analysis in~\cite{LIGOS2iul}.) The
output of the pipeline is a list of candidates which are assigned an SNR
according to Eq.~(\ref{e:combined-snr}). Only three candidates survive in the
final sample.  All these triggers lie in non-playground data and there are no
triple coincident triggers. The two loudest coincident triggers occur when all
three detectors were operating, but the SNR in H1 was too small to cross the
threshold in H2, so they were accepted as coincident triggers according to our
pipeline. The third coincident trigger occurred when only the L1 and H2
detectors were operating.  All three triggers had values of combined SNR very
close to the threshold value of $\rho^{\ast 2} = 61.25$.

A trigger is elevated to the status of an event candidate if the chance
occurrence due to noise is small as measured by the background estimation.
Event candidates are subject to further followup investigations beyond the
level of the automated pipeline to ensure that it is not due to an
instrumental of environmental disturbance. Fig~\ref{f:event-per-s2} shows 
a cumulative histogram of the final coincident triggers versus $\rho^2$
overlayed on the expected background due to accidental coincidences. The
final sample of coincident triggers appear consistent with the expected
background and so we do not believe that they are due to gravitational 
waves. To verify this, further investigation of the three surviving triggers
was performed. There is evidence of transient excess noise in the detectors at
the times of all three triggers, although the origin of this noise could not be 
conclusively identified. The presence of transient noise, the low SNR of the
triggers and their consistency with the expected background rate due to noise
leads us to believe that no gravitational wave signals were detected by the
search.

To determine an upper limit on the event rate we use the \emph{loudest event
statistic}~\cite{loudestGWDAW03} which uses the detection efficiency at the
signal-to-noise ratio of the loudest trigger surviving the pipeline to
determine an upper limit on the rate. The rate of PBH binary inspirals per
Milky Way halo (MWH) is
\begin{equation}
  \mathcal{R}_{90\%} = \frac{2.303+\ln P_b}{T N_H(\rho_\mathrm{max})}
  \ \mathrm{yr}^{-1}\,\mathrm{MWH}^{-1}
\end{equation}
with 90\% confidence. $T$ is the observation time of the search, $N_H$ is the
number of Milky Way halos to which the search is sensitive at the SNR
threshold $\rho^\ast$ of the
loudest trigger $\rho_\mathrm{max}$, and $P_b$ is the probability that all
background triggers have SNR less than $\rho_\mathrm{max}$. This is a
frequentist upper limit on the rate.  For ${\mathcal{R}} >
{\mathcal{R}}_{90\%}$, there is a probability of $90\%$ or greater that at least one
event would be observed with SNR greater than $\rho^\ast$. From the background
analysis, we estimate that $P_b = 0.3\pm0.1$ (statistical error only);
however, for this analysis we omit the background term by setting $P_b = 1$.
This yields a conservative estimate of the upper limit on the rate. 

\begin{figure}
\begin{center}
\includegraphics[width=\linewidth]{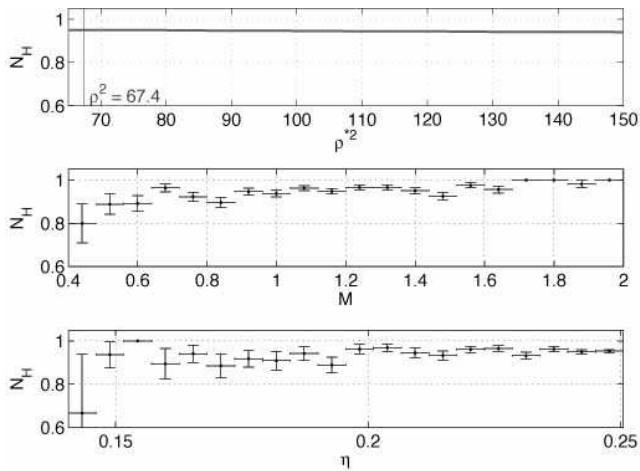}
\end{center}
\caption{\label{f:errors}%
The top panel shows the sensitivity in MW halos $N_H$ of the search to the
target population as a function of the loudest SNR $\rho_\mathrm{max}$. The
largest SNR observed in this analysis was $\rho_\mathrm{max}^2 = 67.4$ meaning
that the search was sensitive to a fraction $N_H = 0.95\ \mathrm{MWH}$ of the
halo. The middle panel shows $N_H$ as a function of
total mass $M = m_1 + m_2$ of the injected signal. The error bars show the
statistical error due to the finite number of injections in the Monte Carlo
simulation. The lower panel shows $N_H$ as a function of the
symmetric mass ratio $\eta = m_1 m_2/ M^2$. We can see that the efficiency is
a weak function of the total mass, as the amplitude of the inspiral signal is
a function of the total mass.  The efficiency of the search does not depend
strongly upon $\eta$.} 
\end{figure}

During the $T = 341$~h~$= 0.0389$~yr of data used in our analysis, the largest SNR
observed was $\rho_\mathrm{max}^2 = 67.4$. The number of Milky Way halos $N_H$ was
computed using a Monte Carlo simulation in which the data was re-analyzed with
simulated inspiral signals drawn from the Milky Way halo population described
by Eq.~(\ref{e:halomodel}). Although we have some sensitivity to the detection
of inspirals with components below this mass, we restrict our upper limit to
the region covered by the template bank $0.2 \le m_1,m_2 \le 1.0\,M_\odot$ by
discarding all injections which have a component mass less than $0.2\,M_\odot$
when computing $N_H$.  Fig.~\ref{f:errors} shows the value of $N_H$ as a
function of the loudest event $\rho^2_\mathrm{max}$.  At $\rho_\mathrm{max}^2
= 67.4$, we find that $N_H = 0.95\ \mathrm{MWH}$.
The various contributions to the error in the measured value of detection
efficiency are described in detail in \cite{LIGOS2iul}. In summary, the
systematic errors are due to uncertainties in the instrumental response,
errors in the waveform due to differences between the true inspiral signal, 
and the finite number of injections in the Monte Carlo simulation. In this
analysis, we neglect errors due to the spatial distribution of the PBH
binaries as studies show that the upper limit is relatively insensitive to the
shape of the Milky Way halo. This is because the maximum range of all three
detectors is greater than $50$~kpc for PBH masses $\ge 0.2\,M_\odot$. The
systematic errors will affect the rate through the measured SNR of the loudest
event.  We can see that the efficiency of the search depends very weakly on
the SNR of the loudest event, again due to the range of the search
compared to the halo radius.  The statistical errors in the Monte Carlo
analysis dominate the errors in $N_H$.  
The combined error due to waveform mismatch and the calibration uncertainty is
found to be $\mathcal{O}(10^{-4})$~MWH. The effects of spin were ignored both
in the population and in the waveforms used to detect inspiral signals.
Estimates based on the work of Apostolatos\cite{Apostolatos:1995} suggest that
the mismatch between the signal from spinning PBHs and our templates will not
significantly affect the upper limit. To be conservative, however, we place an
upper limit only on non-spinning PBHs; we will address this issue quantatively
in future analysis.  Combining the errors in quadrature and assuming the
downward excursion of $N_H$ to be conservative, we obtain an observational
upper limit on the rate of PBH binary coalescence with component masses
$0.2$--$1.0\,M_\odot$ in the Milky Way halo to be
\begin{equation}
\mathcal{R}_{90\%} = 63 \ \mathrm{yr}^{-1}\,\mathrm{MWH}^{-1}.
\end{equation}
\begin{figure}
\begin{center}
\includegraphics[width=\linewidth]{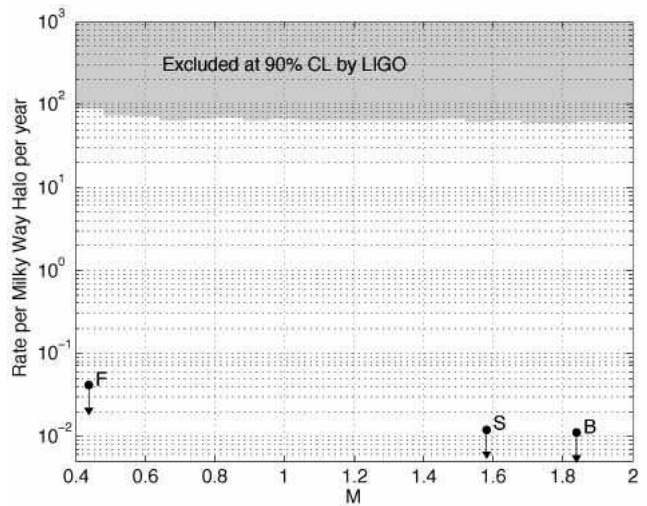}
\end{center}
\caption{\label{f:rate}%
The shaded region shows rates excluded at $90\%$ confidence by the observational
upper limit on PBH binary coalescence presented in this paper as a function of
total mass $M = m_1 + m_2$ of the binary. The three points show the rates
estimated using Eq.~(\ref{e:estrate}) for halo models S ($M = 1.58\,M_\odot$),
F ($M = 0.44\,M_\odot$) and B ($M = 1.84\,M_\odot$) of \cite{Alcock:2000ph}.
}
\end{figure}

By considering numerical simulations of three body PBH interactions in the
early universe Ioka~\emph{et.~al.}~\cite{Ioka:1998nz} obtain a probability
distribution for the formation rate and coalescence time of PBH binaries. This
depends on the PBH mass $m$, which we assume to be the MACHO mass. From this
distribution, we may obtain an estimate of the rate of PBH coalescence at the
present time, given by
\begin{equation}
\mathcal{R} = 1\times 10^{-13} \left(\frac{\mathcal{M}}{M_\odot}\right)
\left(\frac{m}{M_\odot}\right)^{-\frac{32}{37}}\
\mathrm{yr}^{-1}\,\mathrm{MWH}^{-1}
\label{e:estrate}
\end{equation}
where $m$ is the MACHO mass and $\mathcal{M}$ is the mass of the halo in
MACHOs, which is obtained from microlensing observations. These measured
values depend on the halo model used in the analysis of the microlensing
results~\cite{Alcock:1995zx,1996astro.ph..6165A}. The halo model in
Eq.~(\ref{e:halomodel}) corresponds to model S of the MACHO
Collaboration~\cite{Alcock:1995zx}.  The microlensing observations and PBH
formation models assume a $\delta$-function mass distribution, as does the
rate estimate in Eq.~(\ref{e:estrate}). We can see from Fig~\ref{f:errors}
that our detection efficiency is not strongly dependent on the ratio of the
binary masses $\eta$, and so we can marginalize over this parameter to obtain
the rate as a function of total PBH mass $M$, which can be compared with the
predicted rates from microlensing for different halo models. The analysis of
$5.7$~yrs of photometry of $11.9$~million stars in the LMC suggests a MACHO
mass of $m = 0.79^{+0.32}_{-0.24}$ and a halo MACHO mass $\mathcal{M} =
10^{+4}_{-3}\times 10^{10}\, M_\odot$ for halo model S~\cite{Alcock:2000ph}.
Assuming all the MACHOs are PBHs, we obtain the rate estimate $\mathcal{R} =
1.2\times10^{-2} \mathrm{yr}^{-1}\,\mathrm{MWH}^{-1}$, which is three orders
of magnitude lower than our measured rate. Fig~\ref{f:rate} shows a comparison
of the rates predicted using the results of \cite{Alcock:2000ph} for a
standard halo (S), a large halo (B) and small halo (F). Models B and F are
power-law halos~\cite{1993MNRAS.260..191E} and are discussed in detail in
\cite{Alcock:1995zx,1996astro.ph..6165A}. We note that our upper limit is not
strongly dependent on the halo model as all three halos terminate before the
sensitivity of our search is significantly decreased.

Finally we note that the estimated microlensing rate for the standard halo is
lower than that predicted in \cite{Ioka:1998nz}, due to the tighter
constraints placed on the MACHO population by the additional observation time
of \cite{Alcock:2000ph}. At design sensitivity initial LIGO will be able to
see binaries containing $0.5\,M_\odot$ PBHs to
$15$~Mpc~\cite{Nakamura:1997sm}, when averaged over antenna pattern and binary
orientation, suggesting an optimistic rate of several per year. The true rate
may be much lower, or zero if no PBH binaries exist, however the possibility
of detection makes it worthwhile to extend the inspiral search used for binary
neutron stars into the MACHO mass range. In the absence of detection, with
$1$~yr of data at design sensitivity we should be able to place limits on the
rate $\mathcal{R} \sim 10^{-3} \mathrm{yr}^{-1}\,\mathrm{MWH}^{-1}$, assuming
other galaxies have a similar MACHO halo content to our own, and hence
significantly constrain the fraction of MACHOs that may be PBHs.

\acknowledgments
The authors gratefully acknowledge the support of the United States 
National Science Foundation for the construction and operation of the 
LIGO Laboratory and the Particle Physics and Astronomy Research 
Council of the United Kingdom, the Max-Planck-Society and the State of 
Niedersachsen/Germany for support of the construction and operation of 
the GEO600 detector.  The authors also gratefully acknowledge the 
support of the research by these agencies and by the Australian 
Research Council, the Natural Sciences and Engineering Research 
Council of Canada, the Council of Scientific and Industrial Research 
of India, the Department of Science and Technology of India, the 
Spanish Ministerio de Ciencia y Tecnologia, the John Simon Guggenheim 
Foundation, the Leverhulme Trust, the David and Lucile Packard 
Foundation, the Research Corporation, and the Alfred P. Sloan 
Foundation.

\end{document}